# Chapter 1
# Identification of Continuous-Time Dynamical Systems: Neural Network Based Algorithms and Parallel Implementation*

Robert M. Farber[†]    Alan S. Lapedes[†]    Ramiro Rico-Martínez[‡]

Ioannis G. Kevrekidis[‡]

**Abstract**
Time-delay mappings constructed using neural networks have proven successful in performing nonlinear system identification; however, because of their discrete nature, their use in bifurcation analysis of continuous-time systems is limited. This shortcoming can be avoided by embedding the neural networks in a training algorithm that mimics a numerical integrator. Both explicit and implicit integrators can be used. The former case is based on repeated evaluations of the network in a feedforward implementation; the latter relies on a recurrent network implementation. Here the algorithms and their implementation on parallel machines (SIMD and MIMD architectures) are discussed.

## 1 Introduction

Artificial Neural Networks (ANNs) have become a valuable tool for time series processing and nonlinear system identification tasks, particularly in those cases where models based on first principles are not quantitatively predictive, or simply not available. The *ad hoc* models obtained by using experimental measurements to train ANNs can lead to qualitative understanding of the long-term dynamic behavior of the system (its attractors) and their dependence on operating parameters.

For deterministic systems, the future state can in general be obtained as a function of current and previous measurements of a *single* state variable (e.g. [10]). This idea has been successfully exploited to construct time-delay mappings based on highly parallelizable ANN implementations [3]. The ANN constructed typically consists of a linear input layer, one or more (nonlinear) hidden layers and a linear output layer. The current and delayed measurements of the state variable are the inputs to the ANN; additional input neurons can be used to incorporate the dependence on parameters. The output neuron(s) give the prediction of the state variable(s) at a future measurement time.

The dicrete nature of mappings constructed this way prevents them from capturing certain qualitative features of the long term attractors and the bifurcation scenario of the continuous-time system they model (for a discussion see [8]).

An alternative for identification of continuous-time systems, consisting of embedding the ANN in a numerical integrator scheme, has been developed in previous work [1, 7, 8]. Both explicit and implicit integrators can be used; however, though the "explicit integrator" alternative can be readily implemented as a feedforward network, the use of implicit integrators requires *recurrent* network algorithms such as the one suggested in [6] (for another alternative involving recurrent networks see [4, 5]).

---

*This work has been partially supported by DARPA/ONR (N00014-91-J-1850) and a grant by the Exxon Education Foundation.
[†]Theoretical Division. Los Alamos National Laboratory, Los Alamos NM 87545.
[‡]Department of Chemical Engineering. Princeton University, Princeton NJ 08544.

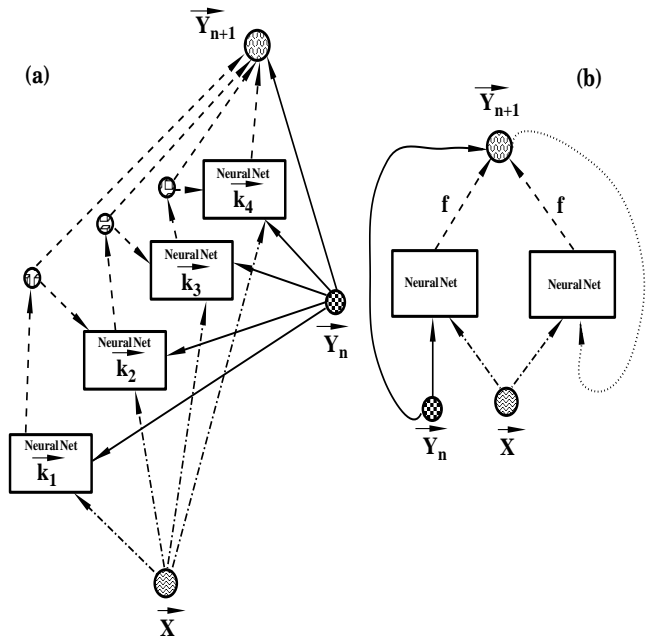

FIG. 1. *ANN schematic for continuous-time identification. (a) A four-layered ANN based on a fourth order Runge-Kutta integrator. (b) ANN embedded in a simple implicit integrator.*

## 2   Continuous-Time Modeling using Neural Networks

Consider the set of autonomous ODEs ($\vec{Y}$: state vector, $\vec{X}$: parameters)

$$\text{(1)} \qquad \dot{\vec{Y}} = \mathbf{f}(\vec{Y}; \vec{X}), \qquad \vec{Y} \epsilon \mathcal{R}^n, \qquad \vec{X} \epsilon \mathcal{R}^p, \qquad \mathbf{f}: \mathcal{R}^n \times \mathcal{R}^p \mapsto \mathcal{R}^n$$

We attempt to construct such a set of ODEs from discrete time data using state measurements only (i.e. without direct numerical evaluation of the time derivatives). The target value is the state of the system at some future time $t+\tau$. We attempt to approximate the unkown right hand side of the ODEs from the experimental knowledge of *integrating* this right hand side (for time $\tau$, with known initial conditions, the state at time t). We assume that the experimental data (the true trajectory of the underlying ODEs) are practically indistinguishable from the result of numerically integrating these ODEs using a simple numerical integration scheme.

For a fourth order Runge-Kutta method, the result $\vec{Y}_{n+1}$ of numerically integrating (1) for fixed values of the operating parameters $\vec{X}$ with initial conditions $\vec{Y}_n$ and a time step of h is given by:

$$\text{(2)} \qquad \vec{Y}_{n+1} = \vec{Y}_n + \frac{1}{6}(\vec{k}_1 + 2\vec{k}_2 + 2\vec{k}_3 + \vec{k}_4), where:$$

$$\vec{k}_1 = h\mathbf{f}(\vec{Y}_n; \vec{X}), \qquad \vec{k}_2 = h\mathbf{f}(\vec{Y}_n + \frac{\vec{k}_1}{2}; \vec{X}), \qquad \vec{k}_3 = h\mathbf{f}(\vec{Y}_n + \frac{\vec{k}_2}{2}; \vec{X}), \qquad \vec{k}_4 = h\mathbf{f}(\vec{Y}_n + \vec{k}_3; \vec{X})$$

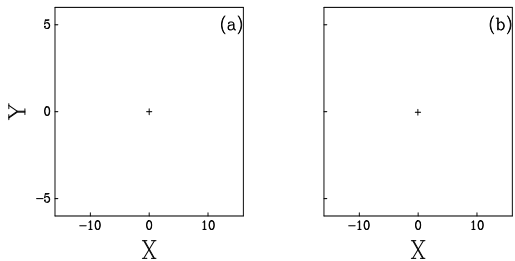

FIG. 2. *(a) Periodic attractor of the Van der Pol oscillator for $\gamma = 1.0$, $\delta = 4.0$ and $\omega = 1.0$. The unstable steady state in the interior of the curve is marked $+$. (b) ANN-based predictions for the attractors of the Van der Pol oscillator shown in (a).*

To approximate the unknown right hand side we use a four-layered ANN schematically described in Figure 1(a). Each ANN "box" (with sigmoidal activation functions in the hidden layers) evaluates $\mathbf{f}(\vec{Y}; \vec{X})$, the right hand side of (1), given $\vec{Y}_n$. The above formulae can be used to derive the weight updating rules for training the overall network. Details of the implementation of this ANN-based procedure and some representative results can be found in [8].

An *explicit* integrator may face problems when the underlying system is "stiff". In this case, the identification procedure can be based on an *implicit* integrator scheme; a simple example is $\vec{Y}_{n+1} = \vec{Y}_n + \frac{h}{2}[\mathbf{f}(\vec{Y}_n; \vec{X}) + \mathbf{f}(\vec{Y}_{n+1}; \vec{X})]$. The resulting overall network is then recurrent [since some of its inputs are equal to its outputs, see figure 1(b)]. The prediction $\vec{Y}_{n+1}$ of the state of the system at time $t + h$ depends on itself. Training this network will require an on-line nonlinear solver, or an algortihm for recurrent networks such as the one suggested in [6]. Further details of the implementation of this algorithm can be found elsewhere [7].

In order to illustrate the capabilities of the methodology described above, we use data from the well-known Van der Pol oscillator: $\dot{x} = -\gamma(y^2 - \delta)x - \omega y, \dot{y} = x$. This system exhibits a stable periodic trajectory for $\gamma = 1.0$, $\delta = 4.0$ and $\omega = 1.0$. Figure 2(a) shows the periodic attractor obtained by numerical integration of the ODEs using a backward difference "stiff" integrator with error control. The unstable steady state (with two positive eigenvalues) at (0,0) is marked $+$ in figure 2(a).

We use data converged on the periodic attractor to construct the training set for the ANN. A total of 240 points are used with "sampling" time 0.1 time units. The ANN "box" embedded in the implicit integrator consists of 2 linear input neurons, two hidden layers with 5 neurons each and a two neuron (linear) output layer. Figure 2(b) shows the predictions given by long-term integration of the right hand side identified by the ANN. In spite of the large sampling time, the ANN correctly captures the character of the oscillations and even predicts the presence of the unstable steady state with good accuracy (both eigenvalues being positive).

## 3 Parallel Implementation

Simulation of recurrent networks is time consuming on a serial machine. Previous work [2] efficiently implemented feedforward networks on a parallel, SIMD, CM2 Connection Machine. The key to the efficiency was the "data parallel" nature of the algorithm [9], spreading the training examples over the processors of the parallel machine: each processor calculates the output of the net for the examples assigned to it. Since one typically has many more examples than synaptic weights, this approach is more efficient on SIMD machines than parallelizing the neurons of a single network over many processors. For recurrent networks, the calculation of the output requires solving the following differential equation for the neuron activity $X_i$ [6]:

$$\dot{X}_i = -\alpha X_i + \beta f(\sum_j W_{ij} X_j) + I_i) \tag{3}$$

where, $W_{ij}$ are the synaptic weights, $I_i$ are external currents, and $\alpha$ and $\beta$ are fixed constants. The solution may be implemented on a SIMD machine if each processor performs the same number of steps of a discrete Euler integration. Alternatively, the calculation may be performed on a MIMD machine, such as the CM5, using any number of algorithms to find the fixed point of (3). The next step of the backpropagation algorithm requires propagation of the "error signal", $Z_i$, through the network. This step requires the solution of a linear set of equations for $Z_i$ [6]: $\sum_r L_{ri} Z_r = J_i$, where $L_{ri}$ is a matrix defined in terms of the synaptic weights and neuron activities, and $J_i$ is defined in terms of the desired fixed points of (3). In the neural literature the solution, $Z_i$, is defined in terms of the fixed point of the linear "network" equation: $\dot{Z}_i = \sum_r L_{ri} Z_r + J_i$. Euler integration of these ODEs may be efficiently performed on SIMD machines for a fixed number of time steps. One can alternatively employ a MIMD implementation to solve the above sets of linear equations. The generalization of our previous "data parallel" SIMD implementations of feedforward neural networks can be easily extended to recurrent networks. Utilization of MIMD machines, such as the CM5, can further increase computational efficiency.

## References

[1] S. R. Chu and R. Shoureshi, *A neural network approach for identification of continuous-time nonlinear dynamic systems*, Proceedings of the 1991 American Control Conference, 1 (1991), pp. 1-5.
[2] R. M. Farber, *Efficiently Modeling Neural Networks On Massively Parallel Computers*, to be published in the proceedings of the NASA workshop on parallel computing Nov. 1991. Los Alamos Technical Report LA-UR-92-3568.
[3] A. S. Lapedes and R. M. Farber, *Nonlinear signal processing using neural networks: prediction and system modeling*, Los Alamos Report LA-UR 87-2662, 1987.
[4] M. Nikolau and V. Hanagandi, *Recurrent neural networks in exact-linearization decoupling control of multivariable nonlinear systems*, 1992 Annual AIChE Meeting, Miami, Fl, Paper 125j (1992).
[5] B. A. Pearlmutter, *Learning state space trajectories in recurrent neural networks*, Proceedings of the 1989 IEEE INNS International Joint Conference in Neural Networks, II (1989), pp. 365-372.
[6] F. J. Pineda, *Generalization of back-propagation to recurrent neural networks*, Phys. Rev. Letters, 59 (1987), pp. 2229-2232.
[7] R. Rico-Martínez and I. G. Kevrekidis, *Continuous time modeling of nonlinear systems: a neural network-based approach*, Proceedings of the 1993 IEEE International Conference on Neural Networks, San Francisco, in press (1993).
[8] R. Rico-Martínez, K. Krischer, I. G. Kevrekidis, M. C. Kube and J. L. Hudson, *Discrete- vs. Continuous-Time Nonlinear Signal Processing of Cu Electrodissolution Data*, Chem. Eng. Comm., 118 (1992), pp. 25-48.
[9] A. Singer, *Implementations of Artificial Neural Networks on the Connection Machine*, Parallel Computing, 14 (1990), pp. 305-315.
[10] F. Takens, *Detecting strange attractors in turbulence*, in Dynamical Systems and Turbulence, Springer, Heidelberg, 1981.